\documentclass[runningheads]{llncs}
\usepackage{graphicx}
\usepackage{amsmath}
\usepackage{arydshln}

\begin{document}
\title{Evaluation of Convolutional Neural Networks for COVID-19 Classification on Chest X-Rays}
%
%
\author{Felipe André Zeiser\inst{1} \and
Cristiano André da Costa\inst{1} \and
Gabriel de Oliveira Ramos \inst{1}\and
Henrique Bohn \inst{1}\and
Ismael Santos \inst{1}\and
Rodrigo da Rosa Righi \inst{1}}
\authorrunning{F. Zeiser et al.}
%
\institute{Graduate Program in Applied Computing\\
    Universidade do Vale do Rio dos Sinos \\
    S\~ao Leopoldo, Brazil\\
\email{\{felipezeiser@edu., cac@, gdoramos@, 	hbohn@edu., ismael@edu., rrrighi@\}unisinos.br }}
\maketitle              
\begin{abstract}
Early identification of patients with COVID-19 is essential to enable adequate treatment and to reduce the burden on the health system. The gold standard for COVID-19 detection is the use of RT-PCR tests. However, due to the high demand for tests, these can take days or even weeks in some regions of Brazil. Thus, an alternative for detecting COVID-19 is the analysis of Digital Chest X-rays (XR). Changes due to COVID-19 can be detected in XR, even in asymptomatic patients. In this context, models based on deep learning have great potential to be used as support systems for diagnosis or as screening tools. In this paper, we propose the evaluation of convolutional neural networks to identify pneumonia due to COVID-19 in XR. The proposed methodology consists of a preprocessing step of the XR, data augmentation, and classification by the convolutional architectures DenseNet121, InceptionResNetV2, InceptionV3, MovileNetV2, ResNet50, and VGG16 pre-trained with the ImageNet dataset. The obtained results demonstrate that the VGG16 architecture obtained superior performance in the classification of XR for the evaluation metrics using the methodology proposed in this article. The obtained results for our methodology demonstrate that the VGG16 architecture presented a superior performance in the classification of XR, with an Accuracy of $85.11\%$, Sensitivity of $85.25\%$, Specificity of $85.16\%$, F1-score of $85.03\%$, and an AUC of $0.9758$.

\end{abstract}
\section{Introduction}

Severe Acute Respiratory Syndrome Coronavirus 2 (SARS-CoV-2) is a new beta-coronavirus first identified in December 2019 in Wuhan Province, China \cite{ANDERSEN2020}. Since the initial outbreak, the number of patients confirmed with COVID-19 has exceeded 178 million in the world. More than 3.85 million people have died as a result of COVID-19 (June 19, 2021) \cite{DONG2020}. These numbers can be even higher due to asymptomatic cases and flawed tracking policies. In Brazil, a study points to seven times more infections than that reported by the authorities \cite{EPICOVID2020}.

SARS-CoV-2 shares $79.6\%$ of the SARS-CoV base pairs genome \cite{ZHOU2020}. Despite its resemblance to the 2002 SARS-CoV virus, SARS-CoV-2 has rapidly spread across the globe challenging health systems \cite{SHEREEN2020}. The burden on healthcare systems is due to the high rates of contagion and how SARS-CoV-2 manifests itself in the infected population \cite{PASCARELLA2020}. According to data from epidemiological analyses, about $20\%$ to $30\%$ of patients are affected by a moderate to severe form of the disease. In addition, approximately $5\%$ to $12\%$ of infected patients require treatment in the Intensive Care Unit (ICU). Of those admitted to ICUs, about $75\%$ are individuals with pre-existing comorbidities or older adults \cite{SEVERE2020}.

In Brazil, the first case was officially notified on February 25, 2020 \cite{BRASIL2020}. Since then, due to the continental proportions of Brazil, several measures to contain and prepare the health system have been carried out in the Federative Units \cite{RAFAEL2020}. Brazil went through two delicate moments for the health system, with hospitals without beds and even the lack of supplies for patients in some regions of the country \cite{TAYLOR2021}.

Currently, reverse polymer transcription chain reaction (RT-PCR) is the test used as the gold standard for the diagnosis of COVID-19 \cite{TANG2020,MARSON2020}. However, due to difficulties in purchasing inputs, increases in the prices of materials and equipment, the lack of laboratories and qualified professionals, and the high demands for RT-PCR tests, the diagnosis can take days or even weeks in some cities in Brazil \cite{MARSON2020}. This delay can directly impact the patient's prognosis and be associated with a greater spread of SARS-CoV-2 in Brazil \cite{CANDIDO2020}.

As alternatives to RT-PCR, radiological exams as Computed Tomography (CT) and Digital Chest Radiography (XR) are being used as valuable tools for the detection and definition of treatment of patients with COVID-19 \cite{COHEN2020}. Studies show equivalent sensitivities to the RT-PCR test using CT images \cite{AI2020,XIE2020}. Lung alterations can be observed even in asymptomatic patients of COVID-19, indicating that the disease can be detected by CT even before the onset of symptoms \cite{LEE2020}.

XR has lower sensitivity rates compared to CT. However, due to some challenges in CT use, health systems adopted XR in the front line of screening and monitoring of COVID-19 \cite{BORGHESI2020}. The main challenges in using CT compared to XR are: (i) exam cost, (ii) increased risk for cross-infection, and (iii) more significant exposure to radiation \cite{WONG2020}. Furthermore, in underdeveloped countries, the infrastructure of health systems generally does not allow RT-PCR tests or the acquisition of CT images for all suspected cases. However, devices for obtaining radiographs are now more widespread and can serve as a fundamental tool in the fight against the epidemic in these countries \cite{COHEN2020}.

In this perspective, artificial intelligence techniques have shown significant results in the processing of large volumes of information in various pathologies and have significant potential in aiding the diagnosis and prognosis of COVID-19 \cite{LAKHANI2017,DEFAUW2018,ZEISER2020,Jatoba2020}. Thus, this work aims to explore the application of Deep Learning (DL) techniques to detect pneumonia caused by COVID-19 through XR images. In particular, our objective is to evaluate the performance and provide a set of pre-trained Convolutional Neural Network (CNN) models for use as diagnostic support systems. The CNN architectures evaluated in this article are: DenseNet121 \cite{HUANG2017}, InceptionResNetV2 \cite{SZEGEDY2017}, InceptionV3 \cite{SZEGEDY2016}, MovileNetV2 \cite{SANDLER2018}, ResNet50V2 \cite{HE2016}, and VGG16 \cite{SIMONYAN2014}.

The study is organized into five sections. In Section~\ref{sec:TrabRelacionados}, we present the most significant related works for the definition of the work. Section~\ref{sec:MatEMet} presents the methodology of the work. Section~\ref{sec:Resultados} details the results. Finally, Section~\ref{sec:Conclusao} presents the conclusions of the work.

\section{Related work}
\label{sec:TrabRelacionados}

Artificial intelligence has made significant advances motivated by large data sets and advances in DL techniques. These advances have allowed the development of systems to aid in analyzing medical images with a precision similar to healthcare specialists. Furthermore, machine learning or data mining techniques extract relevant features and detect or predict pathologies in the images. Therefore, this section describes some relevant works in the literature to detect pneumonia and COVID-19 in XR images.

Since the initial outbreak of COVID-19, studies applying CNNs have been used to detect COVID-19 in XR images. However, at the beginning of the outbreak, the lack of positive XR images was a problem. In \cite{HEMDAM2020}, the performance of seven CNNs using fine-tuning was compared in a set of 50 XR, 25 positive cases, and 25 negative cases for COVID-19. However, the few images used, the lack of a test set, and the learning graphs presented indicate that the models could not generalize the problem. 


In the work proposed by \cite{ABBAS2021}, a CNN-based Decompose, Transfer, and Compose (DeTraC) model is used for COVID-19 XR classification. The model consists of using a pre-trained CNN for the ImageNet set as a feature extractor. Selected features go through Principal Component Analysis (PCA) to reduce dimensionality. These selected characteristics are then classified by a CNN into COVID-19 or not. The proposed model reached an accuracy of 93.1\% and sensitivity of 100\% for the validation set.

DarkCovidNet, a convolutional model based on DarkNet, is used for the COVID-19 detection task in XR \cite{OZTURK2020}. The authors used seventeen DarkNet convolutional layers, achieving an accuracy of 98.08\% for binary classification and 87.02\% for multi-class classification. In \cite{KHAN2020} it is proposed to use the pre-trained CNN Xception for the ImageNet dataset for the classification of images into normal, bacterial, viral, and COVID-19 pneumonia. The study achieved an average accuracy of 87\% and an F1-Score of 93\%.

Another alternative for detecting COVID-19 in XR is detection in levels using VGG-16 \cite{BRUNESE2020}. At the first level, XRs are analyzed to detect pneumonia or not. In the second level, the classification of XRs in COVID-19 or not is performed. Finally, on the third level, the heat maps of the activations for the COVID-19 XRs are presented. The accuracy of the study, according to the authors, is 99\%.

In summary, several recent works have investigated the use of CNNs for the COVID-19 XR classification. However, most of these works were based on small datasets, performed evaluations directly on the validation sets, and only a few performed the multiclass classification. Therefore, these studies lack evidence on their ability to generalize the problem, making it unfeasible to be used as an aid system for the radiologist. Thus, our contribution to these gaps is the proposal to evaluate six convolutional models in a dataset with more than five thousand XR. We evaluated models on a set of XRs not used in training and validation.

\section{Materials and methods}
\label{sec:MatEMet}

\begin{figure}[ht]
    \centering
    \includegraphics[width=1\textwidth]{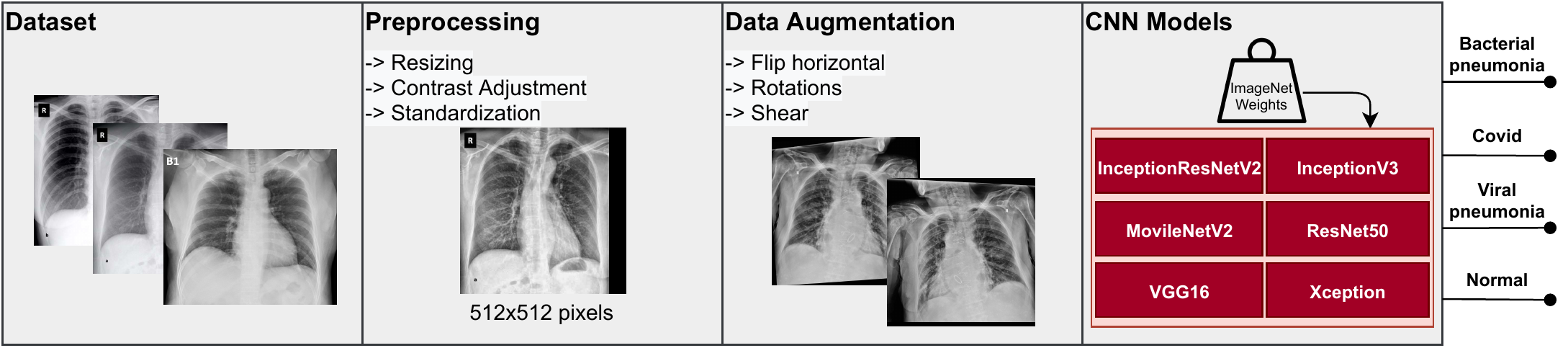}
    \caption{Diagram of the proposed methodology.}
    \label{fig:metodologia}
\end{figure}

An overview of the methodology employed in this work is presented in Fig.~\ref{fig:metodologia}. The methodology is divided into four stages: preprocessing, data augmentation, training, and testing. Image preprocessing consists of image resizing and contrast normalization (Section~\ref{sec:subPreProcessamento}). The data augmentation step describes the methods used to generate synthetic images (Section~\ref{sec:subDataAugmentation}). In the training stage, the CNN models and the parameters used are defined (Section~\ref{sec:subModelosConvolucionais}). Finally, in the test step, we performed the performance evaluation of the CNN models (Section~\ref{sec:subModelosConvolucionais}). 

\subsection{Dataset}

We use XRs from the Curated Dataset for COVID-19 \cite{SAIT2020}. The dataset is a combination of chest XR images from different sources. The dataset has XRs for normal lung, viral pneumonia, bacterial pneumonia, or COVID-19. The authors performed an analysis of all images to avoid having duplicate images. In addition, for the final selection of images, a CNN is used to remove images with noise, such as distortions, cropped, and with annotations. In Tab.~\ref{tab:datasets} we present the number of images used in our work.

\begin{table}[ht]
\centering
\caption{Public dataset of chest XR used in this article.}
\label{tab:datasets}
    \begin{tabular}{c|c}
        \hline
        \textbf{Pathology} & \textbf{Number of images}\\
        \hline
        COVID-19 & 1,281 \\
        \hline
        Normal & 1,300\\
        \hline
        Viral & 1,300 \\
        \hline
        Bacterial & 1,300 \\ 
        \hline
    \end{tabular}
\end{table}

\subsection{Preprocessing}
\label{sec:subPreProcessamento}

In this step, we resize the XRs to $512\times512$ pixels. This size limitation is imposed by available computing power. To avoid distortions in the XR, a proportional reduction was applied to each of the dimensions of the images. We added a black border to complement the size for the dimension that was smaller than $512$ pixels.

\begin{figure}[hbtp!]
\centering
 \begin{tabular}{cc}
        (A) & (B) \\
        \includegraphics[scale=0.205]{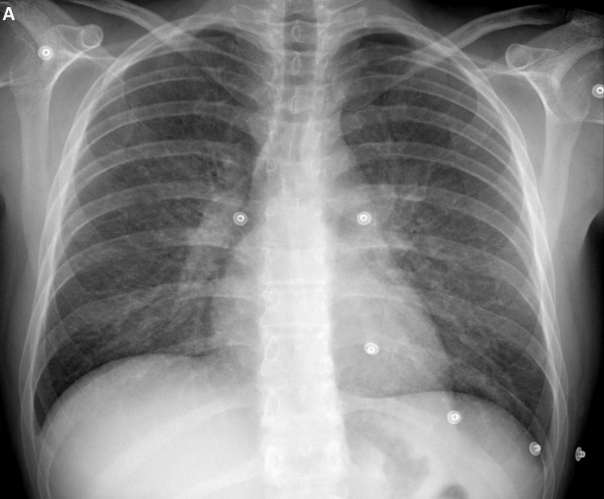} &
        \includegraphics[scale=0.2]{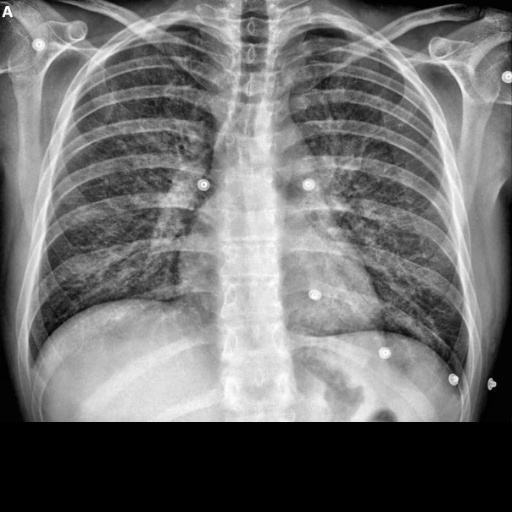}\\
    \end{tabular}
\caption{(A) original XR image with COVID-19; (B) preprocessed image.}
\label{fig:preprocessing}
\end{figure}

As radiographic findings are often low contrast regions, contrast enhancement techniques can be used in XR images. The application of techniques such as Contrast-Limited Adaptive Histogram Equalization (CLAHE) to breast and chest radiography images helped in the generalization of convolutional models and an increase in performance metrics \cite{ZEISER2020,POOCH2020}. Therefore, we use CLAHE to enhance XR images. CLAHE subdivides the image into sub-areas using interpolation between the edges. To avoid noise increase, uses a threshold level of gray, redistributing the pixels above that threshold in the image. CLAHE can be defined by:

\begin{equation}
p = [p_{\text{max}} - p_{\text{min}}]*G(f) + p_{\text{min}}
\label{eqCLAHE}
\end{equation}

\noindent where $p$ is the pixel's new gray level value, the values $p_{\text{max}} $ and $p_{\text{min}} $ are the pixels with the lowest and highest values low in the neighborhood and $G(f)$ corresponds to the cumulative distribution function \cite{CLAHE}. In Fig. \ref{fig:preprocessing} we present an example with the original and preprocessed image.

Finally, we use stratified K-fold cross-validation as a method for evaluating the models, with K=10. With a 10-fold we train the models 10 times for each model. At the end of the training, we calculated the mean and standard deviation of the results for the defined metrics.

\subsection{Data augmentation}
\label{sec:subDataAugmentation}

A technique that helps in the convergence and learning of CNN is the use of data augmentation. Thus, we use the ImageDataGenerator class from Keras to perform on-the-fly data augmentation. We apply a horizontal flip, rotations of up to 20 degrees, and shear in the training set for each fold. We do not apply transformations on XRs in the validation and test set.

\subsection{Convolutional Architectures}
\label{sec:subModelosConvolucionais}

We use six CNN architectures for chest XR classification: DenseNet121 \cite{HUANG2017}, InceptionResNetV2 \cite{SZEGEDY2017}, InceptionV3 \cite{SZEGEDY2016}, MovileNetV2 \cite{SANDLER2018}, ResNet50V2 \cite{HE2016}, and VGG16 \cite{SIMONYAN2014}. We use the pre-trained weights provided by Keras for the ImageNet dataset for each model. This process of initializing pre-trained weights speeds up the convergence process of the models. Tab.~\ref{tab:metaparameters} presents the hyper-parameters used for each of the architectures evaluated in this work.

\begin{table}[ht]
\centering
\caption{Parameters used for each of the CNN architectures.}
\label{tab:metaparameters}
    \resizebox{12cm}{!}{%
        \begin{tabular}{c|c|c|c|c|c}
            \hline
            \textbf{Architecture} & \textbf{Learning Rate} & \textbf{Batch Size} & \textbf{Trainable Params} & \textbf{Non-trainable Params} & \textbf{Depth}\\
            \hline
            DenseNet121 \cite{HUANG2017} &  $5\times 10^{-7}$& $ 16 $ & 11 149 444 & 83 648 & 121 \\
            \hline
            InceptionResNetV2 \cite{SZEGEDY2017} &  $5\times 10^{-7}$ & $ 4 $ & 57 816 420 & 60 544 & 572\\
            \hline
            InceptionV3 \cite{SZEGEDY2016} & $5\times 10^{-7}$ & $ 4 $ & 26 488 228 & 34 432 & 159\\
            \hline
            MovileNetV2 \cite{SANDLER2018} & $5\times 10^{-7}$  & $ 4 $ & 7 468 036 & 34 112 & 88\\
            \hline
            ResNet50V2 \cite{HE2016} & $1\times	 10^{-6}$ & $ 16 $ & 31 909 252 & 45 440 & 50 \\
            \hline
            VGG16 \cite{SIMONYAN2014} & $5\times 10^{-7}$ & $ 16 $ & 48 231 684 & 38 720 & 23 \\
            \hline
    \end{tabular}}
\end{table}

In the training stage, we use categorical cross-entropy as the loss function. The categorical cross-entropy measures the log-likelihood of the probability vector. To optimize the weights of the models, we use the Adam algorithm. At the end of each epoch, we used the validation set to measure the model's accuracy in an independent set and obtain the best training weights.

\section{Results and discussion}
\label{sec:Resultados}

In this section, we present and evaluate the results obtained using the proposed convolutional models.

\subsection{Models training and validation}

\begin{figure}[hbtp!]
\centering
 \begin{tabular}{ccc}
        \includegraphics[scale=0.2]{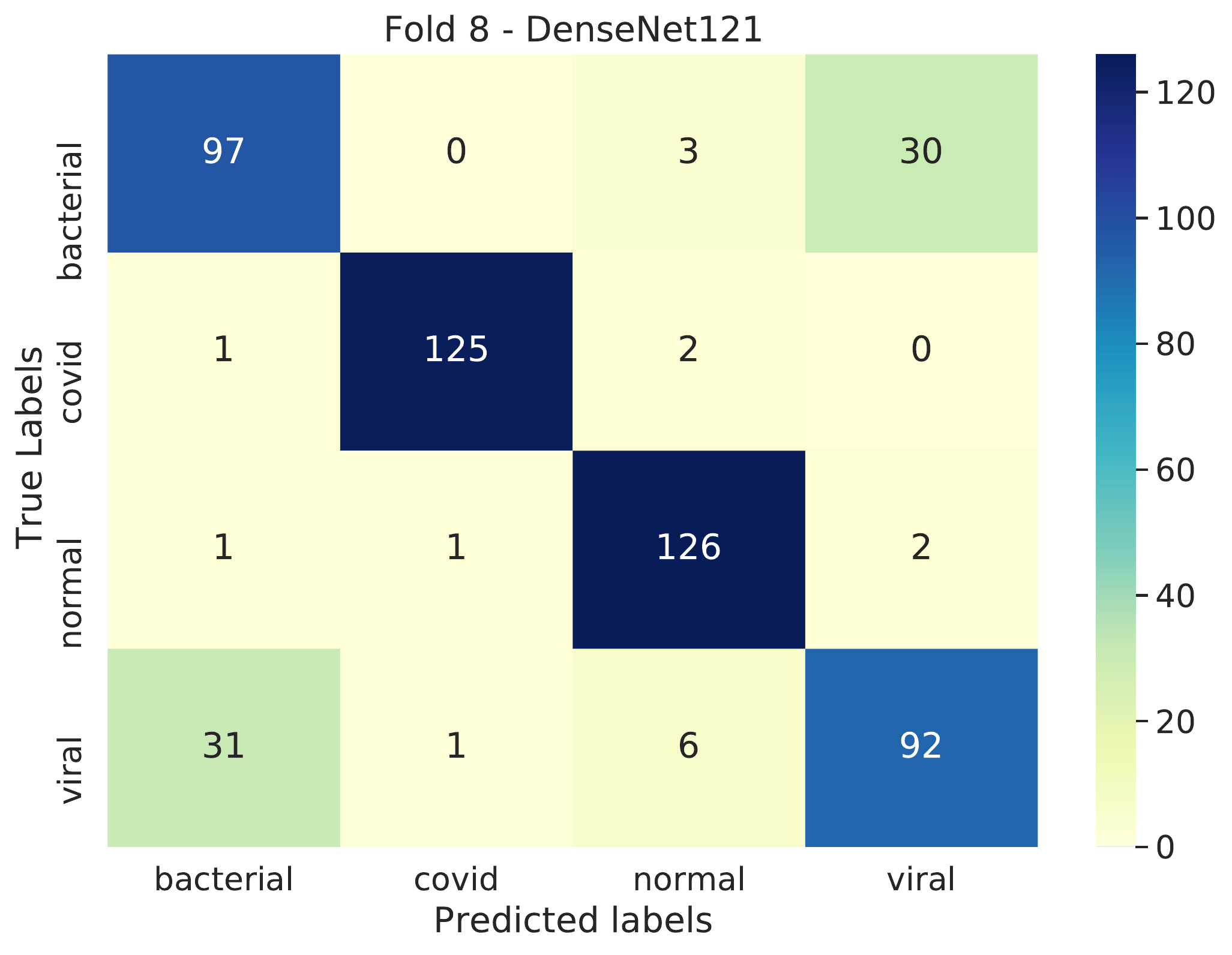} &
        \includegraphics[scale=0.2]{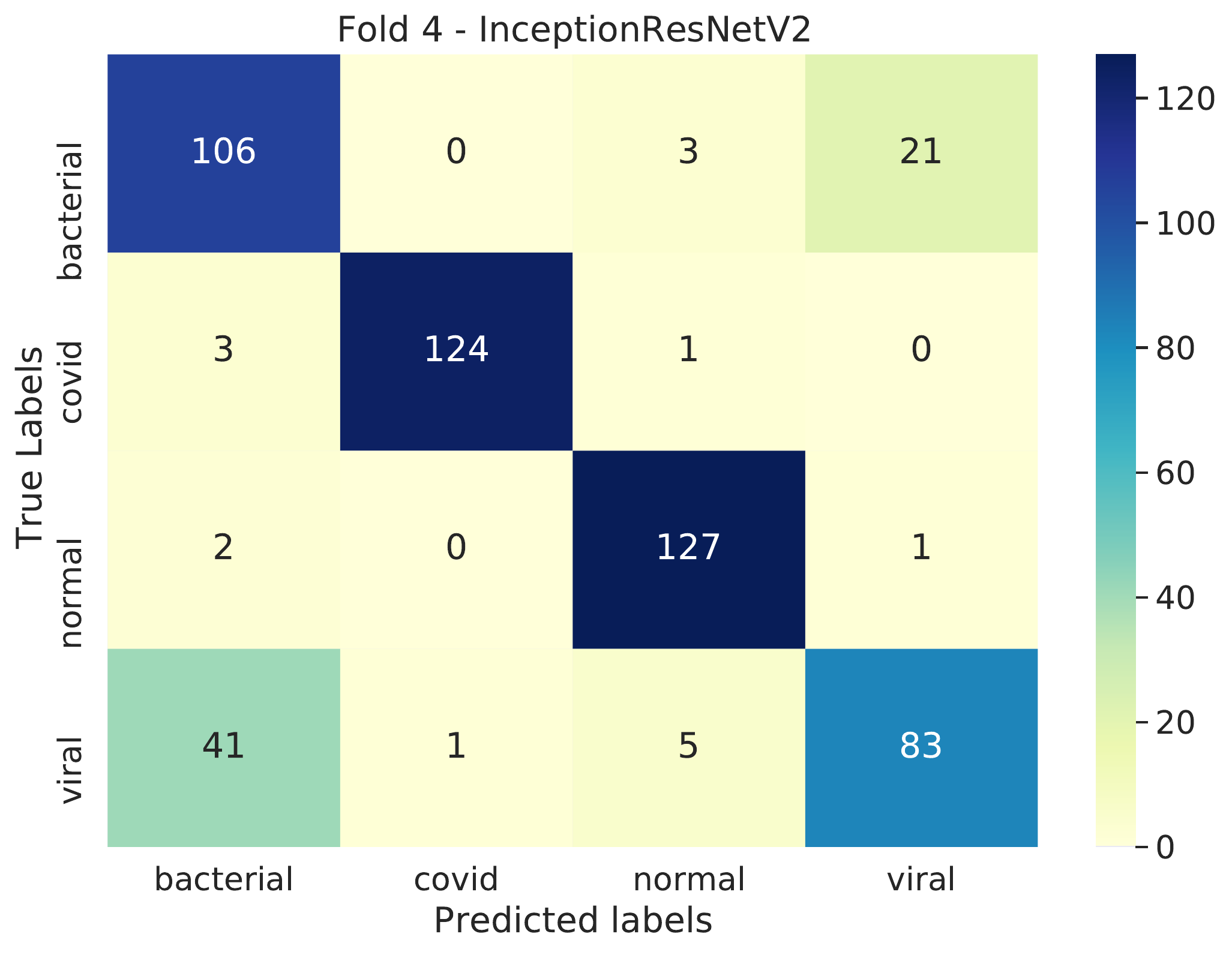} &
        \includegraphics[scale=0.2]{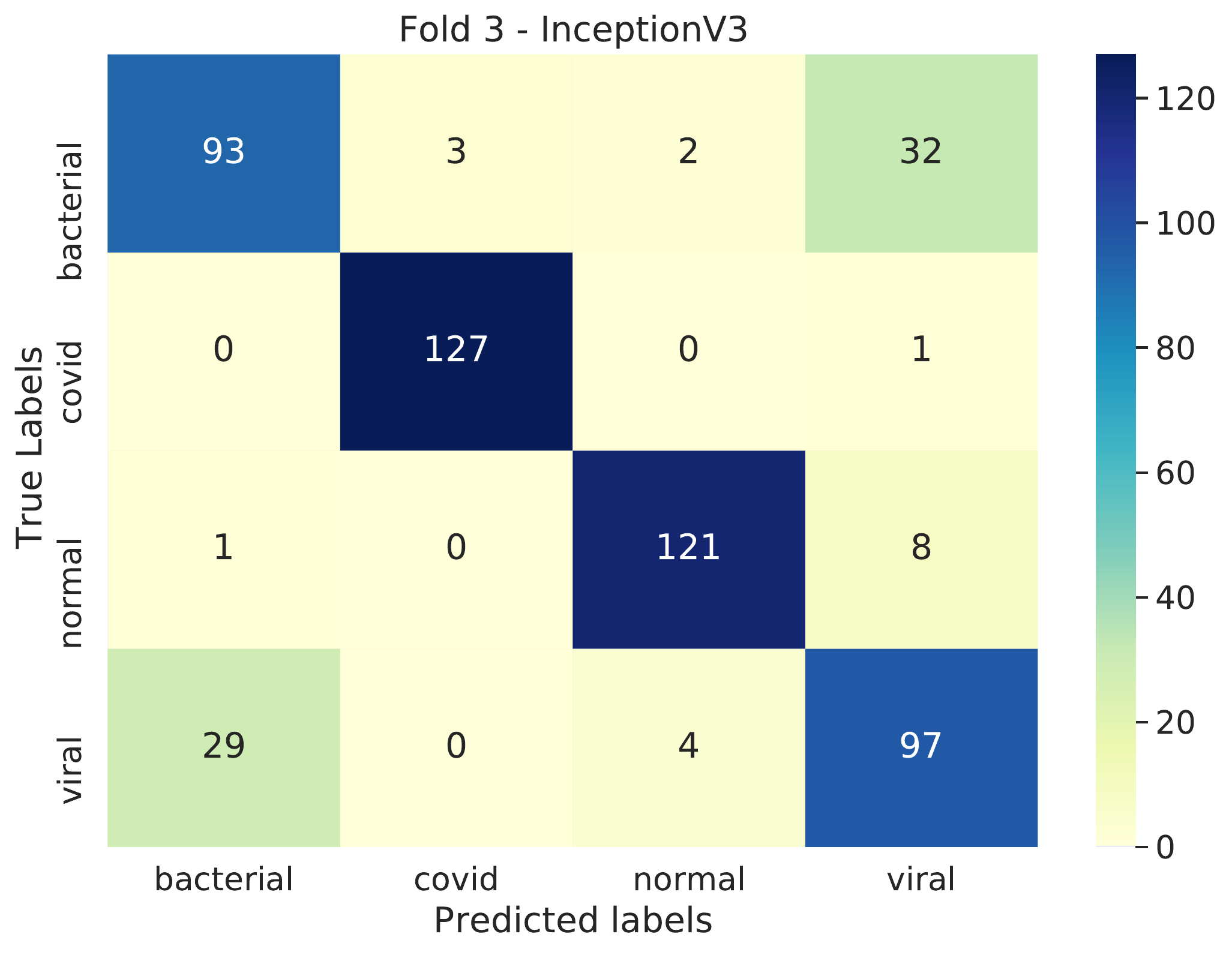}\\
        \includegraphics[scale=0.2]{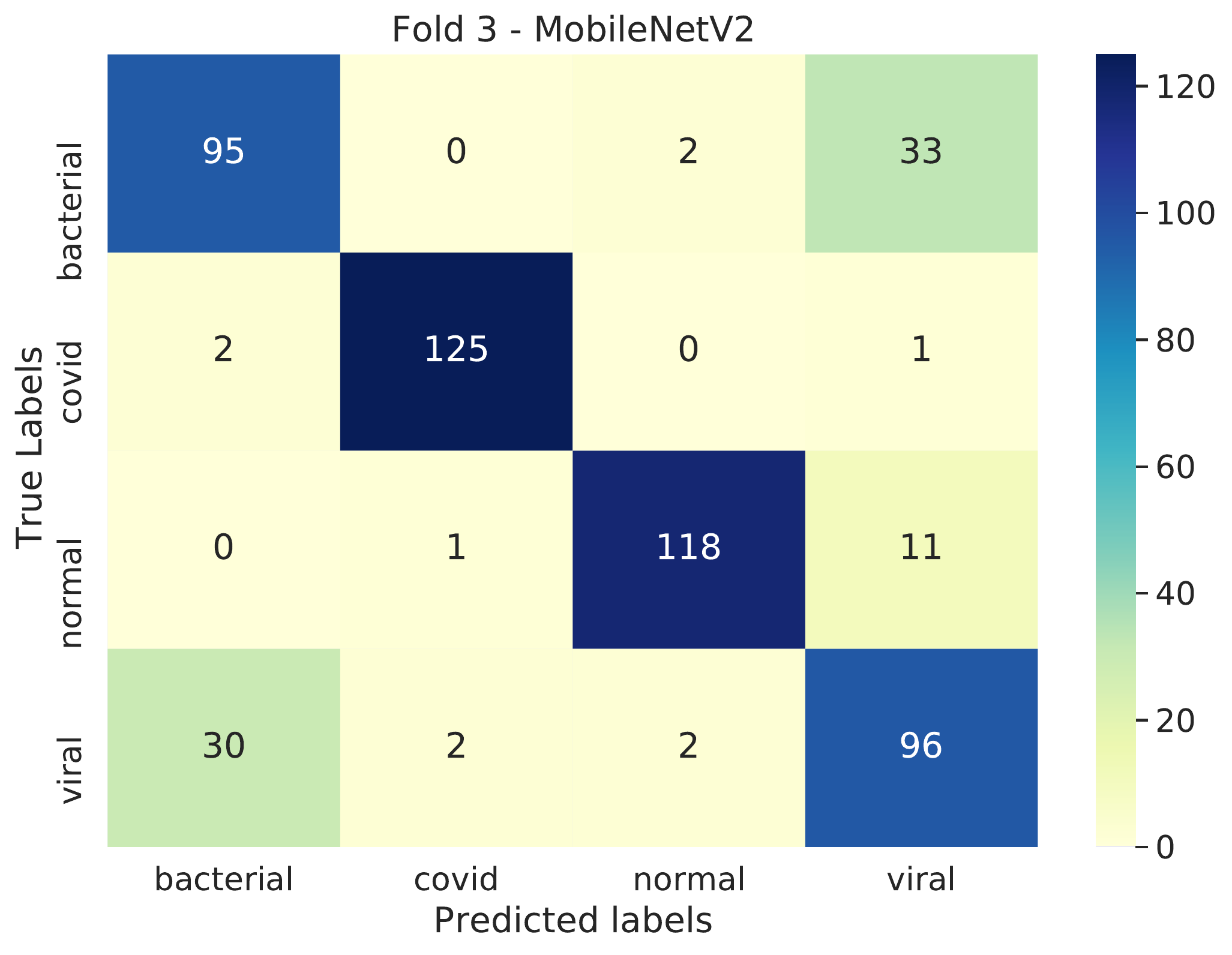} &
        \includegraphics[scale=0.2]{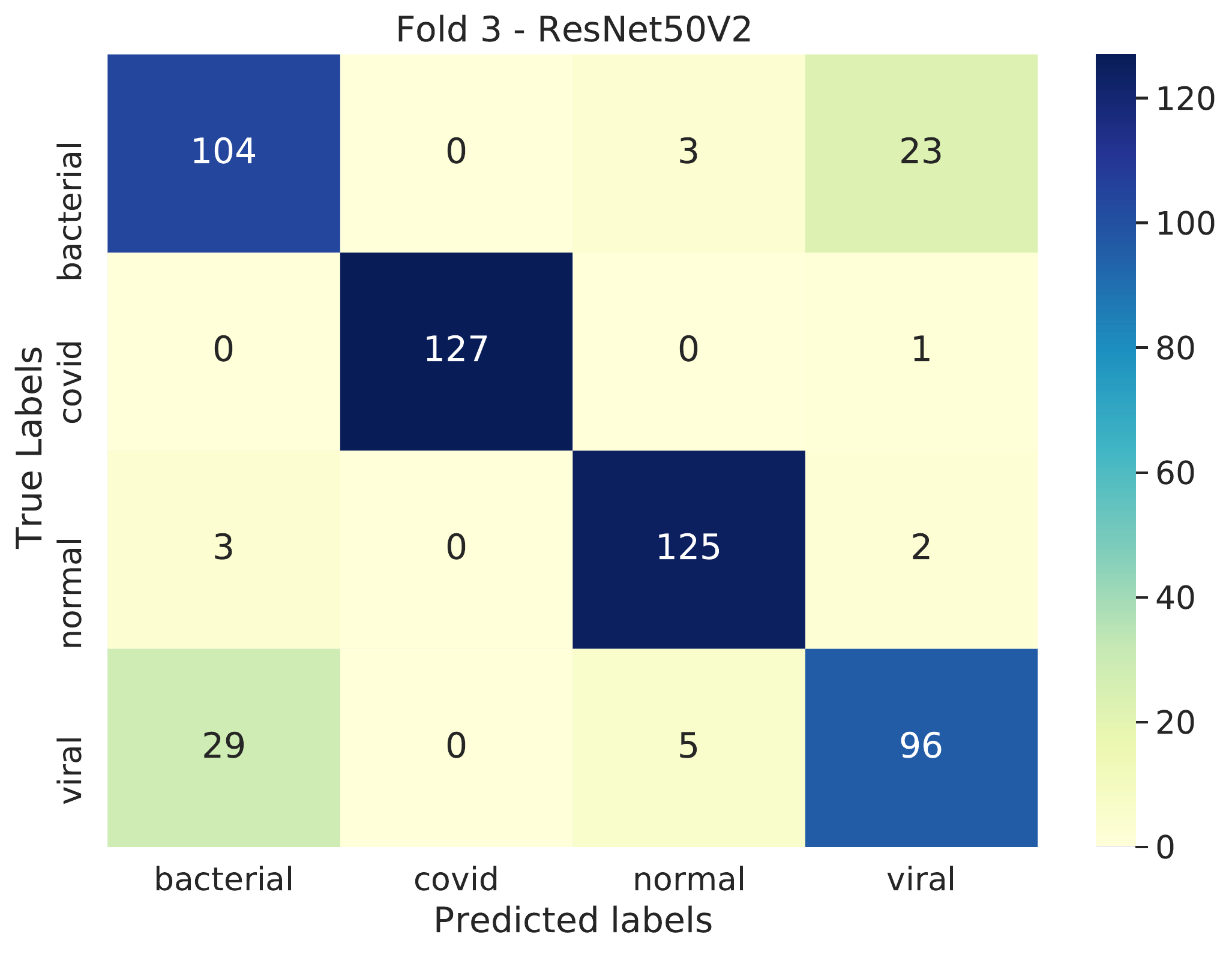} &
        \includegraphics[scale=0.2]{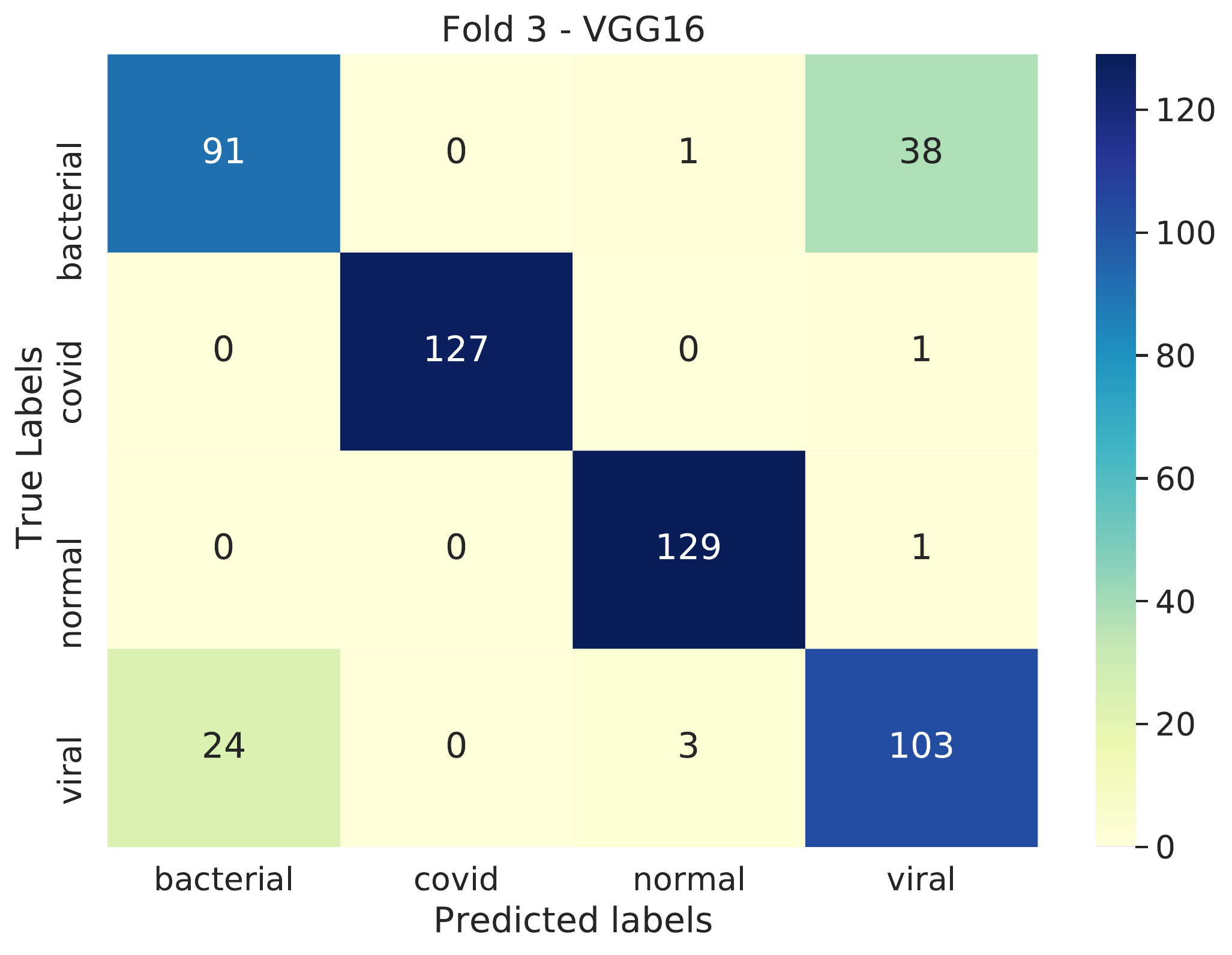}\\
    \end{tabular}
\caption{Confusion matrices of each model for the best test fold set. Dark colors represent a greater number of cases. Light colors represent a smaller amount of cases.}
\label{fig:cm}
\end{figure}

We train the models for 100 epochs for each fold. We evaluate each model at the end of training in the validation set. The choice of the best set of weights was performed automatically based on the error for the validation set. Fig. \ref{fig:cm}, presents the confusion matrices for each of the models.

Analyzing the confusion matrices (Fig. \ref{fig:cm}), we can see that all classifiers could correctly classify the vast majority of cases. We can highlight the tendency to classify cases of viral pneumonia as bacterial and bacterial pneumonia as viral. This trend may indicate that the number of viral and bacterial pneumonia cases was insufficient for an optimized generalization for these two classes. As for the classification of normal cases, the ResNet50V2 model had the highest rate of misclassification. The classification of pneumonia due to COVID-19 showed similar success rates. The highest false-negative rate for COVID-19 was presented by the InceptionResNetV2 model, with 4 cases. The lowest rate of false negatives was presented by the InceptionV3, ResNet-50, and VGG16 models, with 1 cases.

From the confusion matrix, we can calculate the performance metrics of the models \cite{RUUSKA2018}. Tab. \ref{tab:results} presents the values obtained for the evaluation metrics in the test fold based on the confusion matrices presented in Fig. \ref{fig:cm}.

\begin{table}[hbt!]
\centering
\caption{Results for the test fold for each model.}
\label{tab:results}
    \resizebox{\textwidth}{!}{%

    \begin{tabular}{c|c|c|c|c|c}
        \hline
        \textbf{Model} & \textbf{Accuracy} & \textbf{Sensitivity} & \textbf{Specificity} & \textbf{F1-score} & \textbf{AUC} \\
        \hline
        DenseNet121 \cite{HUANG2017} & 81.28$\pm$2.27\% & 81.40$\pm$2.23\% & 81.33$\pm$2.26\% & 81.22$\pm$2.32\% & 0.9620\\
        \hline
        InceptionResNetV2 \cite{SZEGEDY2017} & 84.16$\pm$1.42\% & 83.49$\pm$1.52\% & 84.10$\pm$1.47\% & 84.16$\pm$1.42\% & 0.9707 \\
        \hline
        InceptionV3 \cite{SZEGEDY2016} & 83.14$\pm$1.01\% & 83.34$\pm$1.09\% & 83.20$\pm$1.00\% & 83.22$\pm$1.04\% & 0.9704 \\
        \hline
        MovileNetV2 \cite{SANDLER2018}& 82.04$\pm$1.33\% & 82.55$\pm$1.34\% & 82.10$\pm$1.39\% & 82.21$\pm$1.28\% & 0.9655 \\
        \hline
        ResNet50V2 \cite{HE2016}& 85.08$\pm$1.62\% & \textbf{85.36}$\pm$1.54\% & 85.12$\pm$1.61\% & \textbf{85.06}$\pm$1.60\% & 0.9748 \\
        \hline
        VGG16 \cite{SIMONYAN2014} & \textbf{85.11}$\pm$1.30\% & 85.25$\pm$1.27\% & \textbf{85.16}$\pm$1.30\% & 85.03$\pm$1.42\% & \textbf{0.9758} \\
        \hline
    \end{tabular}}
\end{table}

Analyzing the results, it is clear that there was relative stability in the performance metrics analyzed for each model. For accuracy, the largest standard deviation was $\pm$2.27\%, and the largest difference between the models was 3.83\% (VGG16 and DenseNet121). These results indicate an adequate generalization of each model for detecting pneumonia due to COVID-19. As for sensitivity, which measures the ability to classify positive classes correctly, the models differ by 3.96\%. The maximum variation between models for specificity was 3.83\%.

In general, for the chest XR classification, the ResNet50V2 and VGG16 models showed the best results. This better performance can be associated with the organizations of the models. For ResNet50V2, we can highlight the residual blocks that allow an adaptation of the weights to remove filters that were not useful for the final decision \cite{HE2016}. As for VGG16, the performance may indicate that the classification of XR resources from lower levels, such as more basic forms, is better to differentiate viral pneumonia, bacterial pneumonia, COVID-19, and normal. However, the VGG16 is computationally heavier and requires more training time. Also, VGG16 has a vanishing gradient problem.

\begin{figure}[hbtp!]
\centering
 \begin{tabular}{cc}
        \includegraphics[scale=0.2]{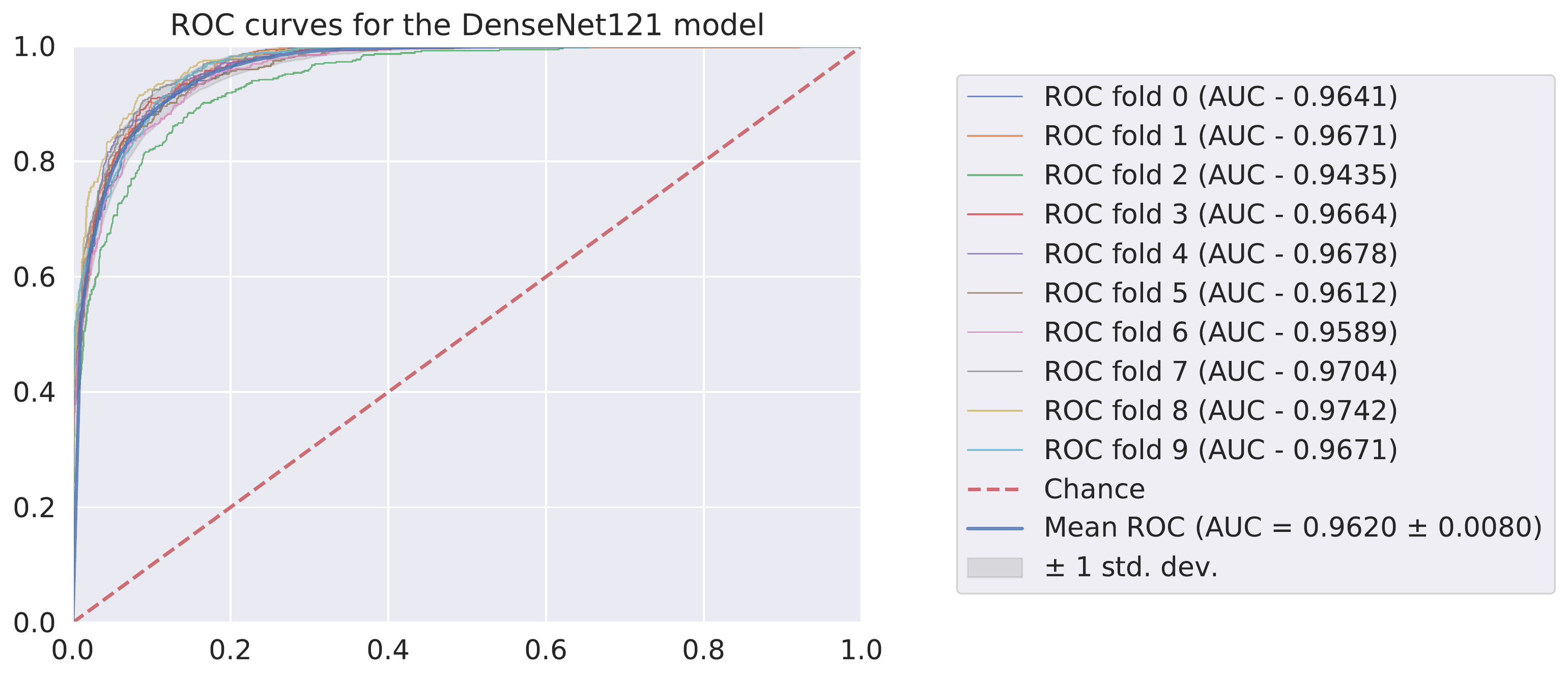} &
        \includegraphics[scale=0.2]{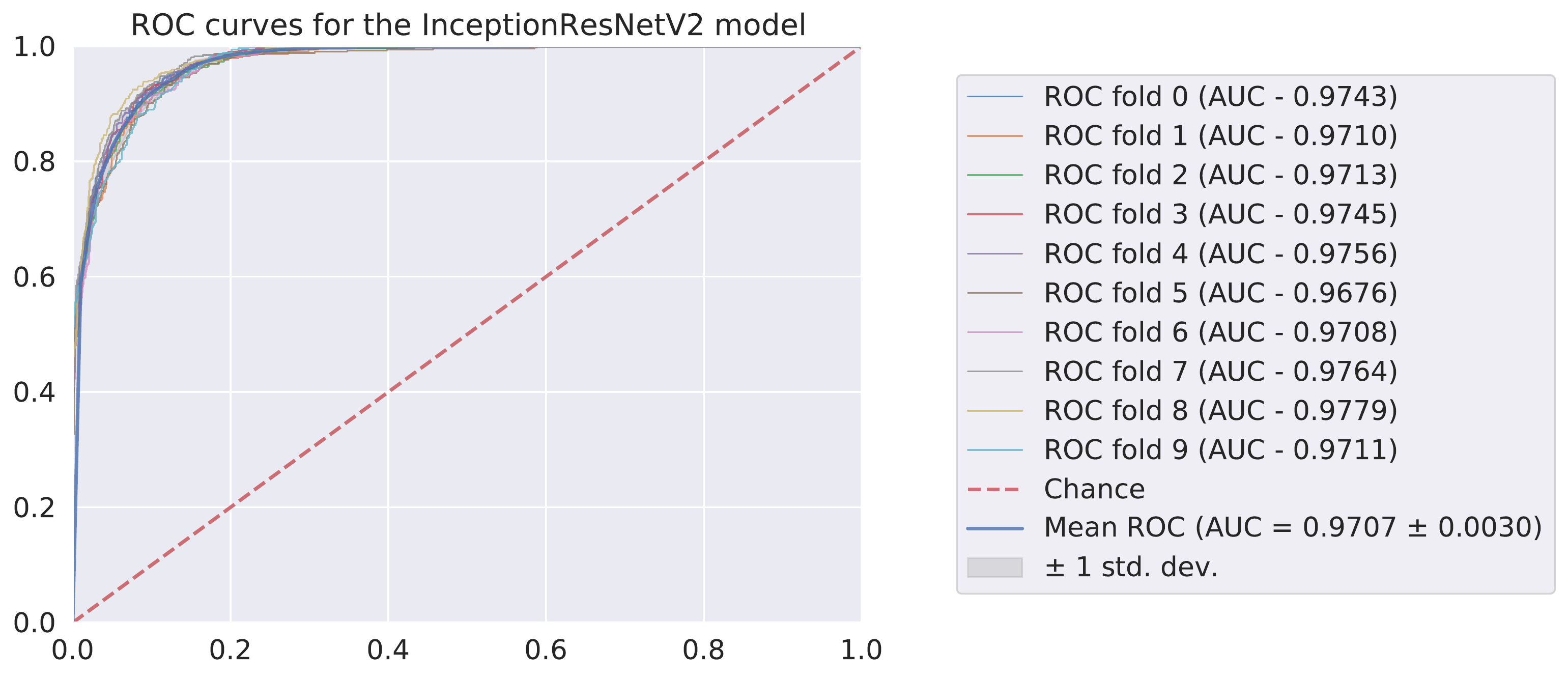} \\
        \includegraphics[scale=0.2]{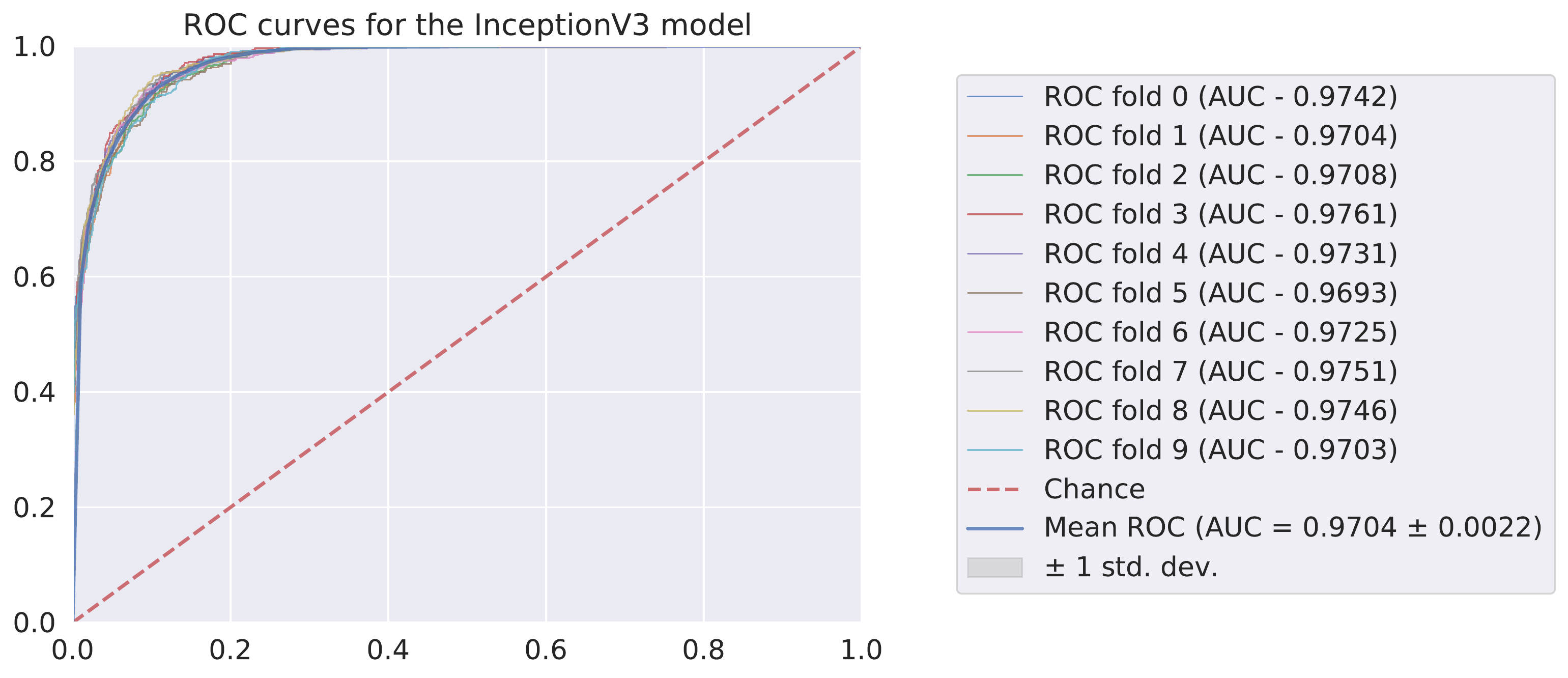} &
        \includegraphics[scale=0.2]{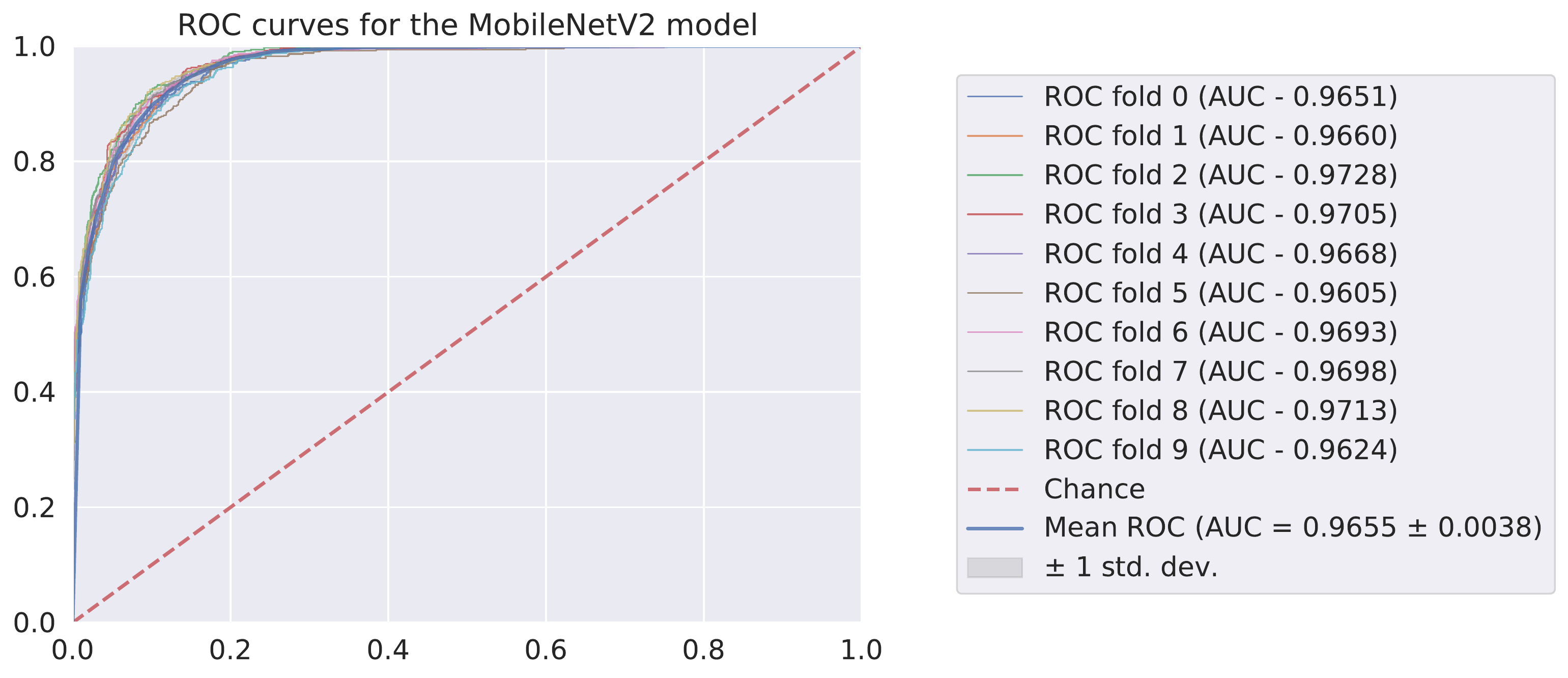} \\
        \includegraphics[scale=0.2]{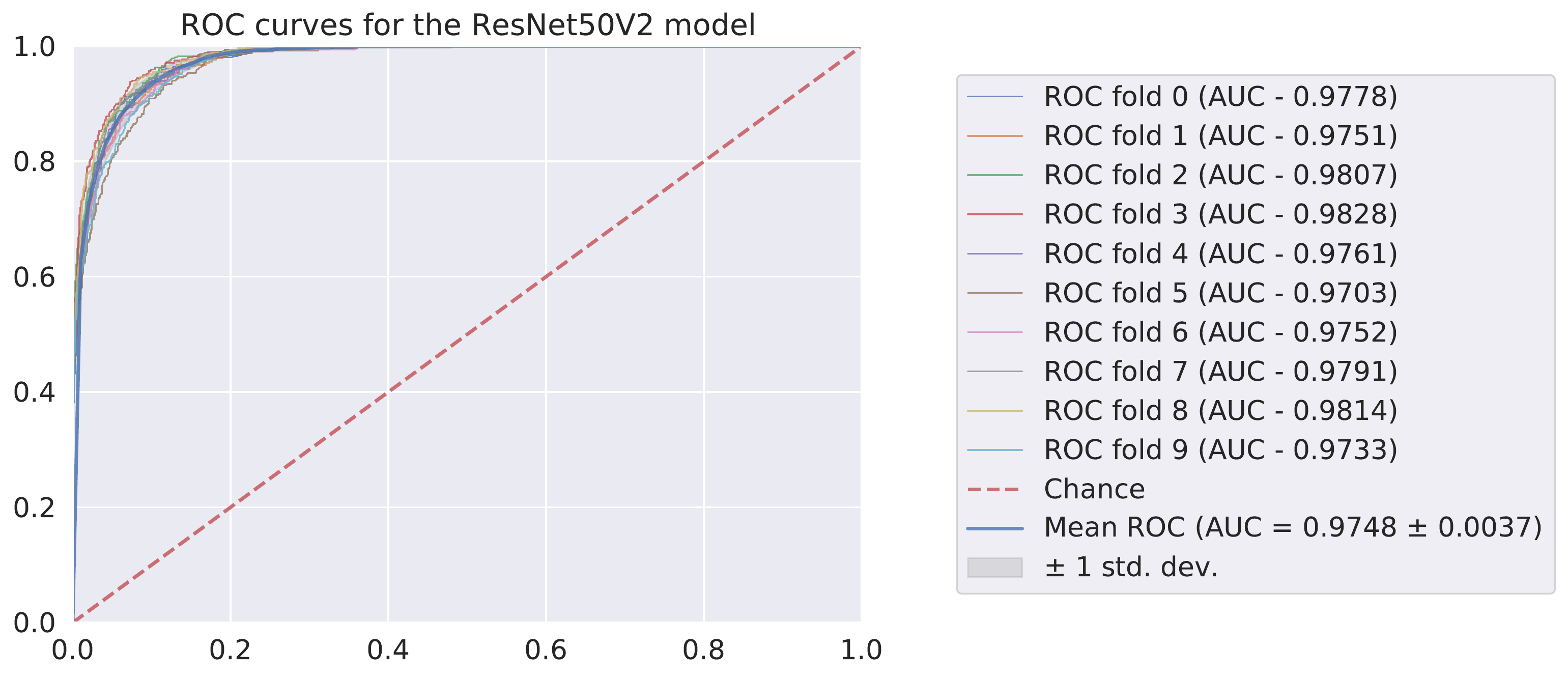} &
        \includegraphics[scale=0.2]{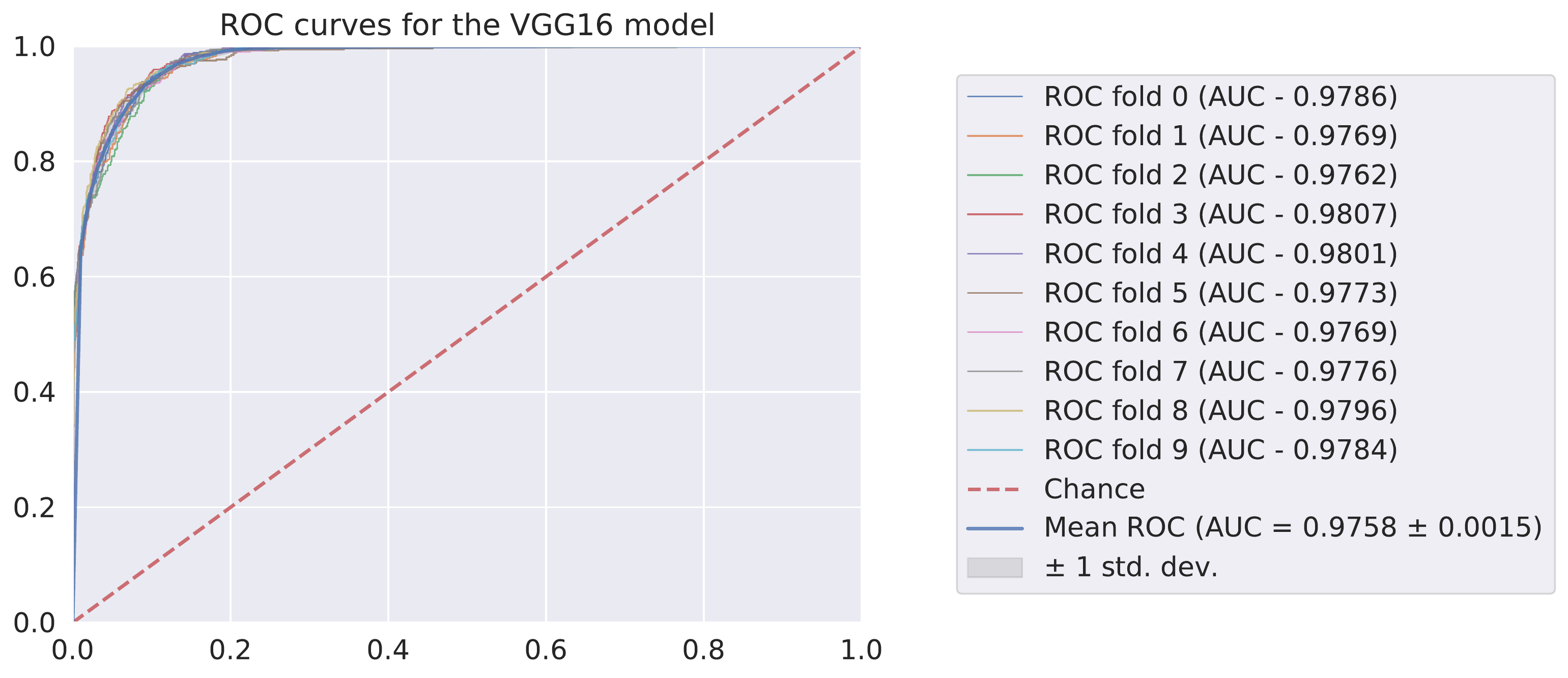}\\
    \end{tabular}
\caption{Classification performance for each fold and model in terms of ROC curves in the test fold.}
\label{fig:roc}
\end{figure}

Fig. \ref{fig:roc} presents the ROC curves for each fold and model in the test fold. When comparing the accuracy, sensitivity, specificity, and F1-score metrics, the Area under the ROC Curve (AUC) showed the greatest stability, with a variation of only $\pm0.80\%$.

\subsection{Comparison with Related Work}

A direct quantitative comparison with related works is difficult. Due to private datasets, methodologies with different image cuts, selecting specific aspects of the datasets, removing some images, and with different classification objectives. In this way, we organized in Table~\ref{tab:comparison_related_work} a quantitative comparison with similar works.

\begin{table}[htb!]
\centering
\caption{Comparison with the related works. \textit{Acc} is the Accuracy, \textit{Sen} is the Sensitivity, \textit{Spe} is Specificity, \textit{F1} is the F1-Score, and \textit{AUC} is the area under the ROC curve.}

\label{tab:comparison_related_work}
\resizebox{\textwidth}{!}{%
    \begin{tabular}{p{0.3\linewidth}|p{0.5\linewidth}|p{0.15\linewidth}|p{0.09\linewidth}|p{0.09\linewidth}|p{0.09\linewidth}|p{0.09\linewidth}|p{0.09\linewidth}}
        \hline
        \textbf{Study} & \textbf{Goal} & \textbf{Dataset size} & \textbf{Acc} & \textbf{Sen} & \textbf{Spe} & \textbf{F1} & \textbf{AUC}\\
        \hline
        Hemdan et al. 2020 \cite{HEMDAM2020} & Comparison of convolutional architectures for binary classification of XR in COVID or not. & 50 & 90.0 & - & - & 91.0 & 0.90  \\
        \hline
        Abbas et al. 2021 \cite{ABBAS2021} & Extract features using CNN with dimensionality reduction using PCA to classify XR into COVID-19, SARS, and normal & 196 & 93.10 & 100 & 85.18 & - & - \\
        \hline
        Ozturk et al. 2020 \cite{OZTURK2020} & Use of DarkNet for multiclass classification of XRs in no findings, pneumonia, and COVID-19 & 1127 & 87.02 & 85.35 & 92.18 & 87.37 & - \\
        \hline
        Khan et al. 2020 \cite{KHAN2020} & Classification of XR into four classes using a pre-trained Xception & 1251 & 89.60 & 89.92 & 96.4 & 89.80 & -\\
        \hline
        Brunese et al. 2020 \cite{BRUNESE2020} & Binary classification of XR in COVID-19 or not at two levels & 6523 & 98.00 & 87.00 & 94.00 & 89.00 & -\\
        \hdashline
        \textit{Our Method} & \textit{Evaluation of CNNs for classification of chest XR into COVID-19, viral pneumonia, bacterial pneumonia and normal} & 5184 & 85.11 & 85.25 & 85.16 & 85.03 & 0.9758\\
        \hline
    \end{tabular}}
\end{table}

In general, the convolutional architectures evaluated in our work achieved performance similar to the works that perform multiclassification \cite{OZTURK2020,KHAN2020,ABBAS2021}. If compared to works that perform binary classification, the difference for accuracy is $\pm$13\%. However, the work with the best accuracy uses a reduced set of positive images for COVID-19 (250 images) out of a total of 6523 \cite{BRUNESE2020}. Thus, the dataset used in the work is not balanced and may indicate a bias in the study. This bias may be associated with a lower sensitivity (87.00\%) of the study compared to other metrics.

The sensitivity was similar to the other studies, except for the work with 100\% sensitivity \cite{ABBAS2021}. However, the size of the dataset used in the study (196 XR images) can influence the results of the method. In addition, the method proposed by \cite{ABBAS2021} is more complex compared to the proposed in our study. In this way, we can highlight that our method use a balanced dataset. We also use stratified K-fold cross-validation for models training. In this way, we can assess the stability of the assessment metrics and the confidence interval. Finally, the trained models and source code is available on GitHub [Hidden ref.].

\section{Conclusion}
\label{sec:Conclusao}

This article compared six convolutional architectures for the detection of pneumonia due to COVID-19 in chest XR images. In order to improve the generalizability of the results, we apply a set of preprocessing techniques. We use several models with pre-trained weights for the ImageNet dataset, and we propose the classification in normal cases, viral pneumonia, bacterial pneumonia, or COVID-19.

The main scientific contribution of this study was the performance comparison of the DenseNet121, InceptionResNetV2, InceptionV3, MovileNetV2, ResNet50, and VGG16 convolutional architectures for the detection pneumonia due to COVID-19. These pre-trained models can serve as a basis for future studies and provide a second opinion to the radiologist during XR analysis.

As future work, we intend to analyze the influence of datasets on the characteristics learning of COVID-19 XR images, analyzing whether CNNs can generalize characteristics for different datasets. Furthermore, we hope to investigate Explainable Artificial Intelligence approaches to convey specialists the features present in the images used to form the diagnostic suggestion. Finally, using multimodal methodologies, for example, using clinical data and images, can be helpful in the transparency of diagnostic recommendations.

\section{ACK}
We thank the anonymous reviewers for their valuable suggestions. The authors would like to thank the Coordination for the Improvement of Higher Education Personnel - CAPES (Financial Code 001), the National Council for Scientific and Technological Development - CNPq (Grant numbers 309537/2020-7), the Research Support Foundation of the State of Rio Grande do Sul - FAPERGS (Grant numbers 08/2020 PPSUS 21/2551-0000118-6), and NVIDIA GPU Grant Program for your support in this work.

\bibliographystyle{splncs04}
\bibliography{references}

\end{document}